\def\ut#1{\mathop{\vtop{\ialign{##\crcr
     $\hfil\displaystyle{#1}\hfil$\crcr\noalign
     {\kern1pt\nointerlineskip}\hbox{$\hfil\sim\hfil$}\crcr
     \noalign{\kern1pt}}}}}

\def\undersymbol#1#2{\mathop{\vtop{\ialign{##\crcr
     $\hfil\displaystyle{#2}\hfil$\crcr\noalign
     {\kern1pt\nointerlineskip}\hbox{$\hfil#1\hfil$}\crcr
     \noalign{\kern1pt}}}}}

\newcommand{\degr}{^0}
\documentstyle[psfig,aaspp4]{article}

\begin{document}

\title{Gamma Ray Astronomy and Baryonic Dark Matter}

\author{F. De Paolis}
\affil{Bartol Research Institute, University of Delaware, Newark, 
DE 19716-4793, USA; depaolis@bartol.udel.edu}

\author{G. Ingrosso}
\affil{Dipartimento di Fisica, Universit\`a di Lecce, CP 193, 73100 Lecce, 
Italy and INFN, Sezione di Lecce, CP 193, 73100 Lecce, Italy; 
ingrosso@le.infn.it}

\author{Ph. Jetzer}
\affil{Paul Scherrer Institut, Laboratory for Astrophysics, 
CH-5232 Villigen PSI, Switzerland
and Institute of Theoretical Physics, University of Zurich, 
Winterthurerstrasse 190, CH-8057 Zurich, Switzerland; jetzer@physik.unizh.ch}

\author{M. Roncadelli}
\affil{INFN, Sezione di Pavia, Via Bassi 6, I-27100, Pavia, Italy; 
roncadelli@pv.infn.it}

\vskip -1cm
\begin{abstract}

Recently, Dixon et al. (\cite{dixon}) have re-analyzed the EGRET data, finding 
a statistically significant diffuse $\gamma$-ray emission from the galactic 
halo. We show that this emission can naturally be explained within a 
previously-proposed model for baryonic dark matter, in which $\gamma$-rays 
are produced through the interaction of high-energy cosmic-ray protons with 
cold $H_2$ clouds clumped into dark clusters - these dark clusters 
supposedly populate the outer galactic halo and can show up in
microlensing observations.
Our estimate for the halo $\gamma$-ray flux turns out to be
in remarkably good agreement 
with the discovery by Dixon et al. (\cite{dixon}).
We also address future prospects to test our predictions.

\end{abstract}
\vfil\eject

\section{Introduction and outlook}
As is well known, the galactic halo chiefly consists of dark matter.
A natural possibility - repeatedly considered in the past
(Silk \cite{silk}, Carr \cite{carr})  - is that
baryonic dark matter makes a substantial contribution.

A few years ago, we
recognized that the Fall \& Rees 
theory for the formation of
globular clusters 
(Fall \& Rees \cite{fall}, 
Kang et al. \cite{kang}, 
Vietri \& Pesce \cite{vietri})
also leads to the existence of dark clusters 
- made of brown dwarfs 
\footnote{
Although we concentrate our attention on brown dwarfs, it should be 
mentioned that red dwarfs as well can be accommodated within the 
considered scenario.} 
and cold self-gravitating  clouds - 
at galactocentric distances $R \ut > 10$ kpc
(De Paolis et al. \cite{depaolis1}-\cite{depaolis4}, 
\cite{depaolisapj}) 
\footnote{Similar models have also been 
proposed by 
Ashman \& Carr (\cite{ac}), Ashman (\cite{ashman}), Fabian and Nulsen 
(\cite{fn1,fn2}), Gerhard \& Silk (\cite{gs}) and 
Kerins (\cite{kerins1,kerins2}).}.
Accordingly, the inner halo is populated by globular clusters whereas the 
outer halo is dominated by dark clusters. 
Contrary to the case of globular clusters, a large amount of residual gas
should remain clumped into the dark clusters, as brown 
dwarfs fail to generate the strong stellar winds which expel the leftover 
gas from globular clusters. Moreover, although the clouds under 
consideration are mainly made of $H_2$, 
we expect them to be surrounded by an atomic layer and a
photo-ionized ``skin'' (De Paolis et al. \cite{depaolisapj}).
We stress that the presence of cold self-gravitating clouds in the 
halo is a characteristic feature of the model in question. 
Remarkably enough, quite 
recently it has been pointed out (Walker \& Wardle \cite{ww}) that cold 
self-gravitating clouds of the considered kind naturally explain the 
``extreme scattering events'' associated with compact radio quasars 
(Fiedler et al. \cite{fiedler}).

Our proposal was also motivated by the discovery
of microlensing events towards the LMC (Alcock et al. 
\cite{alcock1,alcock2}, 
Aubourg et al. \cite{aubourg}). 
The first-year data were manifestly
consistent with the assumption that MACHOs are brown dwarfs
even within the standard (isothermal) galactic model. Unfortunately,
the present situation is much less clear. 
An option is that the halo resembles more closely a maximal disk rather than 
an isothermal sphere, in which case MACHOs can still be brown dwarfs
\footnote{
It should be kept in mind that a large fraction of MACHOs
(up to $\sim 50\%$ in mass) can be binary systems, thereby counting as 
twice more massive objects (De Paolis et al. \cite{depaolismnras}).}
(see also Binney \cite{binney}).
A more intriguing possibility has recently been suggested by 
Kerins \& Evans (\cite{ke}) 
and naturally fits within our model. As the initial mass 
function evidently changes with the galactic distance $R$, it can well 
happen that brown dwarfs dominate the halo mass density without 
however dominating the optical depth for microlensing \footnote{Notice that 
also the considered clouds can contribute to microlensing events (Draine
\cite{draine}).}.

A few months ago, Dixon et al. (\cite{dixon}) have re-analyzed the EGRET data 
concerning the diffuse $\gamma$-ray flux with a wavelet-based technique. 
After subtraction of the isotropic extragalactic component and of the 
expected contribution from the Milky Way, they find a statistically 
significant diffuse emission from the galactic halo. At high-galactic 
latitude, the integrated halo flux above 1 GeV turns out to be 
$\simeq 10^{-7}-10^{-6}$ $\gamma$ cm$^{-2}$ s$^{-1}$ sr$^{-1}$, 
which is slightly less than the 
diffuse extragalactic flux (Sreekumar et al. \cite{sreekumar}).

Our aim is to show that the diffuse $\gamma$-ray
emission from the galactic halo discovered 
by Dixon et al. (\cite{dixon}) can naturally
be explained within the considered model.
Basically, the idea is that cosmic-ray (CR) protons in the
galactic halo scatter on 
halo clouds, thereby producing the observed $\gamma$-ray flux.

\section{Cosmic ray confinement in the galactic halo}

Unfortunately, neither theory nor observation allow nowadays to
make sharp statements about the propagation of CRs in the galactic
halo \footnote {We stress 
that - contrary to the practice used in the CR community - 
by halo we mean the (almost)
spherical galactic component which extends beyond 
$\sim$ 10 kpc.}.
Therefore, the only possibility to get some insight into this issue 
rests upon the extrapolation from the knowledge of CR propagation in the 
disk. Actually, this strategy looks sensible, since the leading effect is CR 
scattering on inhomogeneities of the magnetic field over scales
from $10^2$ pc down to less than $10^{-6}$ pc
(Berezinsky et al. \cite{berezinskii})
and - according to our model - inhomogeneities 
of this kind are expected to be present in the halo,
because of the existence of gas clouds - with a photo-ionized ``skin'' -
clumped into dark clusters
\footnote{
Indeed, typical values of the dark cluster radius are
$\sim 10$ pc, whereas typical values of the cloud radius
are $\sim 10^{-5}$ pc (De Paolis et al. \cite{depaolisapj}).}.

As is well known, CRs up to energies of $\sim 10^6$ GeV 
are confined in the galactic disk for $\sim 10^7$ yr. 
CRs escaping from the disk will further diffuse in the galactic
halo, where they can be retained for a long time,
owing to the scattering on the above-mentioned
small inhomogeneities of the halo magnetic field
\footnote{A similar idea has been proposed with a somewhat different 
motivation by Wdowczyk \& Wolfendale (\cite{wolf}).}.

Indirect evidence that CRs are in fact trapped in a low-density
halo has recently been reported. For example, Simpson and Connell
(\cite{simpson}) argue that, based on measurements of isotopic abundances
of the cosmic ratio $^{26}$Al/$^{27}$Al, the CR lifetimes are perhaps a 
factor of four larger than previously thought, thereby implying that CRs 
traverse an average density smaller than that of the galactic disk. 
 
A straightforward extension of the diffusion model 
(Berezinsky et al. \cite{berezinskii})
implies that the CR escape time $\tau_{\rm esc}^{~H}$ from the halo 
(of size $R_H$ much larger than the disk half-thickness) is given by
\begin{equation}
\tau_{\rm esc}^{~H} \simeq \frac{R_H^2}{3D_H(E)}~,
\label{tau}
\end{equation}
where $D_H(E)$ is the diffusion coefficient.

We recall that - for CR propagation in the disk - the diffusion coefficient is
$D(E) \simeq D_0~(E/7~ GeV)^{0.3}$ cm$^2$ s$^{-1}$ in the ultra-relativistic 
regime, whereas it reads
$D(E) \simeq D_0 \simeq 3 \times 10^{28}$ cm$^2$ s$^{-1}$ in 
the non-relativistic regime (see Berezinsky et al. \cite{berezinskii}).
As a matter of fact, radio observations in clusters of galaxies
yield for the corresponding diffusion constant $D_0$ 
a value similar to that found in the galactic
disk (Schlickeiser, Sievers \& Thiemann \cite{sst})
\footnote{Moreover, we note that average magnetic field 
values in galactic halos 
are expected to be close to those of galaxy clusters, 
i.e.  in the range 0.1 - 1 $\mu$G (Hillas \cite{hillas}).}.
So, it looks plausible that a similar value for $D_0$ 
also holds on intermediate scale lengths, namely within the galactic halo.
In the lack of any further information on the energy-dependence of $D_H(E)$,
we assume the same dependence as that established for the disk.
Hence, from eq. (\ref{tau}) - with $R_H \sim 100$ kpc - 
we find that for energies $E \ut < 10^3$ GeV the escape
time of CRs from the halo is greater than the age of the Galaxy
$t_0 \simeq 10^{10}$ yr
(notice that below the ultra-relativistic regime 
$\tau_{\rm esc}^{~H}$ gets even longer). 
As a consequence - since the CR flux 
scales like $E^{-2.7}$ (see next Section) - protons with $E \ut < 10^3$
GeV turn out to give the leading contribution to the CR flux.

We are now in position to evaluate the CR energy density in the halo, getting
\begin{equation}
\rho_{CR}^{~H} \simeq \frac{3 t_0 L_G }{4 \pi R_H^3} \simeq 0.12
~~~~~{\rm eV~cm^{-3}}~,
\label{hcrd}
\end{equation}
where 
$L_G \simeq 10^{41}$ erg s$^{-1}$ is the galactic CR luminosity.
Notice, for comparison, that $\rho_{CR}^{~H}$ turns out to be about
one tenth of the disk value (Gaisser \cite{gaisser}).

We remark that we have taken specific realistic values for the various
parameters entering the above equations, in order to make a quantitative 
estimate.
However, somewhat different values can be used. For instance,
$R_H$ may range up to $\sim 200$ kpc (Bahcall, Lubin \& Dorman
\cite{bld}),
whereas $D_0$ might be as large as $\simeq 10^{29}$ cm$^2$ s$^{-1}$
consistently with our assumptions. Moreover, $L_G$ can be as large
as $3 \times 10^{41}$ erg s$^{-1}$ (V\"olk, Aharonian \& Breitschwerdt
\cite{volk}). It is easy to see 
that these variations do not substantially affect our previous conclusions.

\section{Gamma-ray emission from halo clouds}

We proceed to estimate the total $\gamma$-ray flux produced by halo 
clouds clumped into the dark clusters through the interaction
with high-energy CR protons. 
CR protons scatter on cloud protons giving rise (in particular) to pions, 
which subsequently decay into photons.
We
expect negligible high-energy ($\geq$ 100 MeV) $\gamma$-ray photon 
absorption outside the clouds,
since the mean free path is orders of magnitudes larger than the 
halo size. 
 
An essential ingredient is the knowledge of both $\rho^{~H}_{CR}$ 
and the CR spectrum $\Phi_{CR}^{~H}(E)$ in the galactic halo. 
According to the discussion in the previous Section, 
we take $\rho^{~H}_{CR}\simeq 0.12$ eV cm$^{-3}$.
As far as $\Phi_{CR}^{~H}(E)$ is concerned, 
we adopt the following power-law
\begin{equation}
\Phi_{CR}^{~H}(E) \simeq \frac{A}{{\rm GeV}} 
\left(\frac{E}{{\rm GeV}}\right)^{-\alpha}~~~
{\rm particles~cm^{-2}~s^{-1}~sr^{-1}}~,
\label{eqno:42}
\end{equation}
where the constant $A$ is fixed by the requirement that the integrated
energy flux (in the range $1~ {\rm GeV} \leq E \leq 10^3~ {\rm GeV}$)
agrees with the above value of $\rho^{~H}_{CR}$.
The choice of $\alpha$ is nontrivial. As an orientation, the observed
spectrum of primary CRs on Earth entails 
$\alpha \simeq 2.7$. However, this conclusion cannot be extrapolated
to an arbitrary region in the halo (and in the disk), since 
$\alpha$ crucially depends on the diffusion processes undergone by
CRs. For instance, the best fit to EGRET data in the disk towards 
the galactic centre yields $\alpha \simeq 2.45$ (Mori \cite{mori}),
thereby showing that $\alpha$ increases as a consequence of diffusion.
In the lack of any direct information, we conservatively
take $\alpha \simeq 2.7$ 
even in the halo, but in the Table we report for comparison some results
for different
values of $\alpha$.
As can be seen, the flux does not vary substantially.

Let us next turn our attention to the evaluation of the $\gamma$-ray flux
produced in halo clouds 
through the reactions $pp \rightarrow \pi^0 \rightarrow \gamma
\gamma$. The source function $q_{\gamma}(>E_{\gamma},\rho,l,b)$ 
- yielding the photon number density at distance
$\rho$ from Earth in the direction $(l,b)$ with energy $>E_{\gamma}$ - is 
\begin{equation}
q_{\gamma}(>E_{\gamma},\rho,l,b)=
\frac{4\pi}{m_p}\rho_{H_2}(\rho,l,b) 
\int_{E_p(E_{\gamma})}^{\infty}
d\bar{E}_p~ \Phi^{~H}_{CR}(\bar{E}_p)~ 
\sigma_{in}(p_{lab}) <n_{\gamma}(\bar{E}_p)>~~~
{\rm \gamma~cm^{-3}~s^{-1}}~,
\label{eqno:49}
\end{equation}
where the lower integration limit $E_p(E_{\gamma})$ is the minimal proton 
energy necessary to produce a photon with energy $>E_{\gamma}$,
$\sigma_{in}(p_{lab})$ is the inelastic pion production cross-section,
$n_{\gamma}(\bar{E}_p)$ is the photon multiplicity
\footnote{
For the inclusive cross-section of the reaction 
$pp \rightarrow  \pi^{0}  \rightarrow \gamma \gamma$ 
we employ the parameterization given by Dermer (\cite{dermer}).}
and $\rho_{H_2}(\rho,l,b)$ is the halo gas density profile
\footnote{
As it would be exceedingly difficult to keep track of the
clumpiness of the actual gas distribution, we assume that its
density goes like the dark matter density - anyhow, the very low 
angular resolution of $\gamma$-ray detectors would not permit to
distinguish between the two situations.},
which reads
\footnote{
As usual, we use the coordinate transformation
$x = -\rho \cos b \cos l +R_0$,
$y = -\rho \cos b \sin l$ and 
$z =  \rho \sin b$,
where $R_0 = 8.5$ kpc is our galactocentric distance.}
\begin{equation}
\rho_{H_2}(x,y,z) = f~ {\rho_0 (q)} ~ \frac{a^2+R_0^2}{a^2+x^2+y^2+(z/q)^2}~,
\label{eqno:29}
\end{equation}
for $\sqrt{ x^2+y^2+z^2/q^2} > R_{min}$, otherwise it vanishes. Here
$R_{min} \simeq 10$ kpc is the minimal galactocentric distance of 
the dark clusters 
in the galactic halo,
$f$ denotes the fraction of halo dark matter in the form of gas,
$\rho_0(q)$ is the local dark matter density, 
$a = 5.6$ kpc is the core radius and $q$ parametrizes
the halo flattening. For the standard spherical halo model 
$\rho_0(q=1) \simeq 0.3$ GeV cm$^{-3}$, whereas it turns out that e.g. 
$\rho_0(q=0.5) \simeq 0.6$ GeV cm$^{-3}$.

Because $dV=\rho^2 d\rho d\Omega$, it follows that the 
$\gamma$-ray flux per unit solid angle produced in halo clouds and
observed on Earth from the direction $(l,b)$ is
\begin{equation}
\Phi_{\gamma}^{~ \rm DM}
(>E_{\gamma},l,b)=\frac{1}{4\pi} 
\int^{\rho_2(l,b)}_{\rho_1(l,b)} d\rho~ q_{\gamma}(>E_{\gamma},\rho,l,b)
~~~\gamma~{\rm cm^{-2}~s^{-1}~sr^{-1}}~, 
\label{eqno:51}
\end{equation}
where typical 
values of $\rho_1$ and $\rho_2$ are 10 kpc and 100 kpc, respectively.

\section{Discussion and conclusions}

Our main result - which follows directly from eq. (\ref{eqno:51}) -
are maps for the intensity distribution of the $\gamma$-ray emission from 
baryonic dark matter in the halo. In order to make the discussion definite,
we take $f \simeq 0.5$. 

In Figure 1 we show the contour plots in the first quadrant of the 
sky ($0\degr \le l \le 180\degr$, $0\degr \le b \le 90\degr$) for 
the $\gamma$-ray flux  at energy $E_{\gamma}>1$ GeV 
$\Phi_{\gamma}^{~\rm DM} (> 1 {~\rm GeV})$.
Corresponding contour plots for $E_{\gamma}>0.1$ GeV are identical,
up to an overall constant factor equal to 8.74 (again, this follows
from eq. (\ref{eqno:51})).

Figure 1a refers to a spherical halo, whereas Figure 1b pertains to a $q=0.5$ flattened halo.
We see that - regardless of the adopted value for $q$ -
at high-galactic 
latitude $\Phi_{\gamma}^{~\rm DM}(>1{\rm ~GeV})$ lies in the range
$\simeq 6-8 \times 10^{-7}$ $\gamma$ cm$^{-2}$ s$^{-1}$ sr$^{-1}$. 
However, the shape of the contour lines strongly depends on
the flatness parameter. 
Indeed, for $q \ut > 0.9$ there are two contour lines (for 
each flux value) approximately symmetric with respect to $l=90\degr$
(see Figure 1a). 
On the other hand, for 
$q \ut < 0.9$ there is a single contour line (for each value of the flux) 
which varies much less with the longitude (see Figure 1b). 

As we can see from the Table and the Figures, the predicted 
value for the $\gamma$-ray flux at high-galactic latitude
is very close to that found by Dixon et al. (\cite{dixon}).
This conclusion holds almost irrespectively of the flatness parameter.
Moreover, the comparison of the overall shape of the contour lines in our 
Figures 1a and 1b with the corresponding ones in Figure 3 of 
Dixon et al. (\cite{dixon}) entails that models with flatness parameter
$q \ut < 0.8$ are in better agreement with data, thereby 
implying that most likely 
the halo dark matter is not spherically distributed.

We remark that eq. (\ref{eqno:51}) yields
$\Phi_{\gamma}^{~\rm DM} (> 0.1 {\rm~GeV}) \simeq 5.9 \times
10^{-6}$ $~\gamma$ s$^{-1}$ cm$^{-2}$ sr$^{-1}$ at high-galactic latitude
(for a spherical halo). 
This value is roughly 40\% of the diffuse extragalactic $\gamma$-ray emission
of $1.45 \pm 0.05 \times 10^{-5}$ 
$~\gamma$ s$^{-1}$ cm$^{-2}$ sr$^{-1}$ found by the EGRET team
(Sreekumar et al. \cite{sreekumar}). 
So, our result supports the conclusion of Dixon et al. (\cite{dixon})
that the halo $\gamma$-ray emission is a relevant fraction of
the isotropic diffuse flux also 
for $E_{\gamma} > 0.1{\rm~GeV}.$

Before closing this Letter, we would like to briefly
address the crucial question whether the newly discovered $\gamma$-ray
halo emission really calls for a dark matter source. For, one might suspect
that a nonstandard inverse-Compton $\gamma$-ray production 
mechanism could explain the data
(owing to the large uncertainties both in the electron hight scale
and in the electron injection spectral index). However, this seems 
not to be the case. 
Basically, the inverse-Compton contour lines decrease much more rapidly
than the observed ones. Hence, it would be impossible
to explain in this manner the $\gamma$-ray flux found 
by Dixon et al. (\cite{dixon}) while still correctly accounting for the 
observed disk emission (Sreekumar et al. \cite{sreekumar}). 
A more detailed account of this topic will be presented elsewhere.

In conclusion, we feel that - in spite of the various uncertainties -
the remarkably good agreement
between theory and experiment makes our model for halo dark matter worth 
further consideration. 
In particular, the next generation of $\gamma$-ray satellites like
AGILE and GLAST can test our prediction, thanks to the higher sensitivity
and the better angular resolution. In this respect, it might be interesting
to measure whether there is an enhancement in the $\gamma$-ray
flux towards the nearby M31 galaxy, since we expect a similar mechanism for the
$\gamma$-ray production to hold in its halo as well.

\acknowledgments
The work of FDP is supported by an INFN grant.
We would like to thank G. Bignami, P. Caraveo, D. Dixon,
T. Gaisser, M. Gibilisco,
G. Kanbach, T. Stanev, A. Strong and M. Tavani for useful discussions.

\begin{center}
\begin{figure*} 
\vspace{16  cm}
\includegraphics{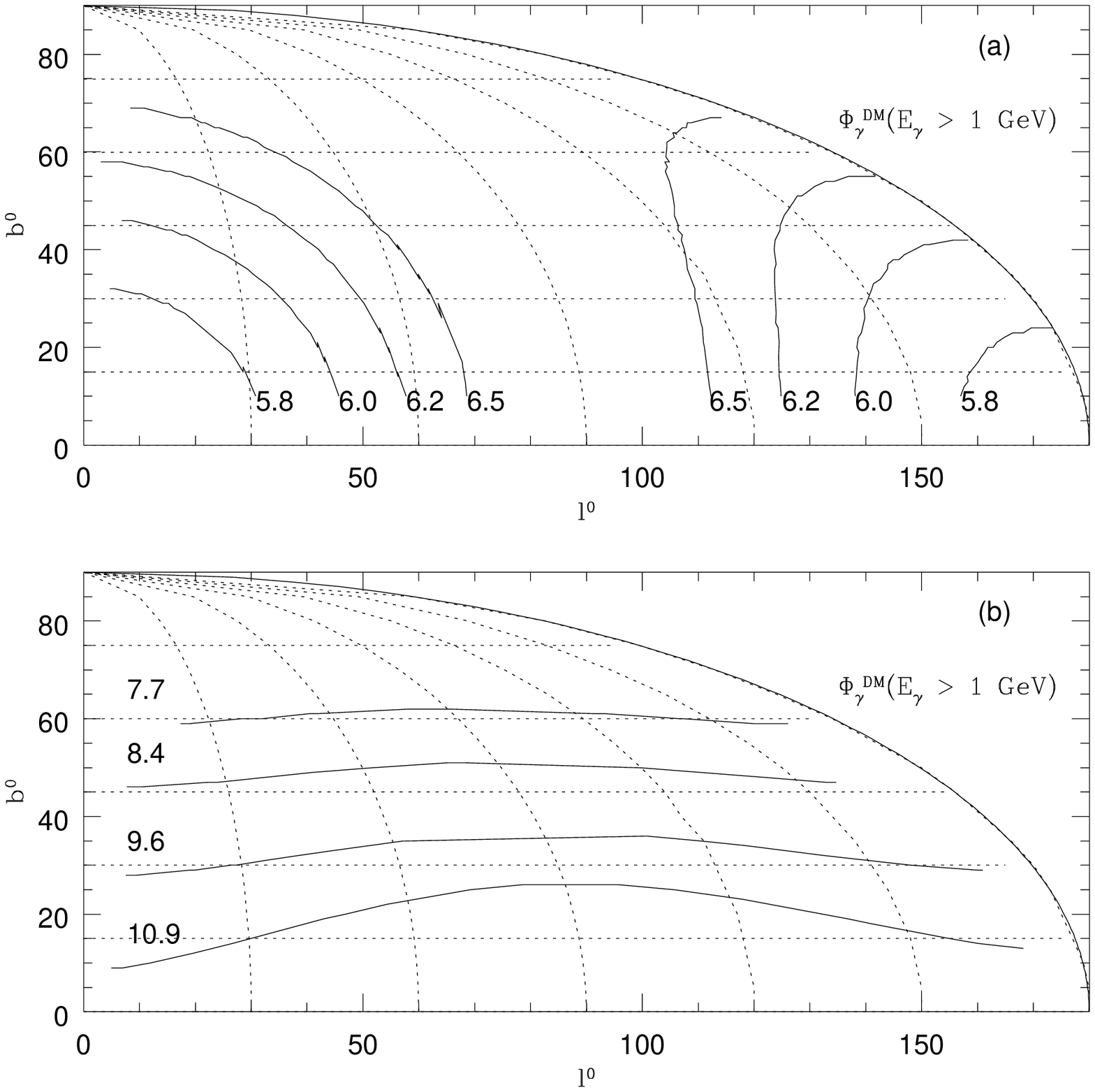} 
\caption{Contour values for the halo $\gamma$-ray flux  
for $E_{\gamma} > 1$ GeV are shown for the indicated values 
in units of $10^{-7}$ $~\gamma$ cm$^{-2}$ s$^{-1}$ sr$^{-1}$,
in the two cases: (a) spherical halo, (b) $q=0.5$ flattened halo.}  
\label{fig2}
\end{figure*}
\end{center}

\def\e{$\pm$}
\begin{deluxetable}{lcccccccccc}
\footnotesize
\tablewidth{0pt}
\tablenum{1}
\tablecaption{
Halo $\gamma$-ray intensity 
at high-galactic latitude for a spherical halo 
evaluated for 
$R_{min}= 10$ and 15 kpc at 
energies above 0.1 GeV and 1 GeV for different values of the CR spectral 
index $\alpha$ (see eq. (\ref{eqno:42})).} 
\tablehead{$R_{min}   $ & $E_{\gamma}$ & $\alpha$  &  $\Phi_{\gamma}^{~\rm DM} (b=90\degr)$
\nl
 (kpc)       &        (GeV) &           &  ($\gamma$  cm$^{-2}$ s$^{-1}$ sr$^{-1}$ ) 
}
\startdata
\hline
$10$ & $0.1$ &  2.45&   $6.2 \times 10^{-6} $ \\
     &       &  2.70&   $5.9 \times 10^{-6} $ \\
     &       &  3.00&   $4.9 \times 10^{-6} $ \\
\hline
$10$ & $ 1$  &  2.45&   $1.1 \times 10^{-6} $ \\
     &       &  2.70&   $6.7 \times 10^{-7} $ \\
     &       &  3.00&   $3.3 \times 10^{-7} $ \\
\hline
\hline
$15$ & $0.1$ &  2.45&   $3.7 \times 10^{-6} $ \\
     &       &  2.70&   $3.5 \times 10^{-6} $ \\
     &       &  3.00&   $2.9 \times 10^{-6} $ \\
\hline
$15$ & $1$   &  2.45&   $6.5 \times 10^{-7} $ \\
     &       &  2.70&   $4.0 \times 10^{-7} $ \\
     &       &  3.00&   $1.9 \times 10^{-7} $ \\
\hline
\enddata
\end{deluxetable}
\normalsize


\begin{thebibliography}{}
\bibitem[1993]{alcock1}
Alcock, C. et al. 1993, Nat 365, 621
\bibitem[1997]{alcock2}
Alcock, C. et al. 1997, ApJ 486, 697
\bibitem[1990]{ashman} 
Ashman, K. M. 1990, MNRAS 247, 662
\bibitem[1988]{ac}
Ashman, K. M. \& Carr, B. J. 1988, MNRAS 234, 219
\bibitem[1993]{aubourg}
Aubourg, E. et al. 1993, Nat 365, 623
\bibitem[1995]{bld}
Bahcall, N. A., Lubin, L. M. \& Dorman, V. 1995, ApJ 447, L81
\bibitem[1990]{berezinskii}
Berezinskii, V. S. et al., 1990,
{\it Astrophysics of cosmic rays} (North-Holland, Amsterdam)
\bibitem[1998]{binney} Binney, J., astro-ph 9809097
\bibitem[1994]{carr}
Carr, B. 1994, Ann. Rev. Astron. Astrophys. 32, 531
\bibitem[1995a]{depaolis1}
De Paolis, F., Ingrosso, G., Jetzer, Ph. \& Roncadelli, M. 1995a, Phys. Rev. 
Lett. 74, 14 
\bibitem[1995b]{depaolis2}
De Paolis, F., Ingrosso, G., Jetzer, Ph. \& Roncadelli, M. 1995b, A\&A 
295, 567
\bibitem[1995c]{depaolis3}
De Paolis, F., Ingrosso, G., Jetzer, Ph., Qadir, A. \&
Roncadelli, M. 1995c, A\&A 299, 647
\bibitem[1995d]{depaolis4}
De Paolis, F., Ingrosso, G., Jetzer, Ph. \& Roncadelli, M. 1995d,
Comments on Astrophys. 18, 87
\bibitem[1998a]{depaolisapj}
De Paolis, F., Ingrosso, G., Jetzer, Ph. \& Roncadelli, M. 1998a,
ApJ 500, 59
\bibitem[1998b]{depaolismnras}
De Paolis, F., Ingrosso, G., Jetzer, Ph. \& Roncadelli, M. 1998b,
MNRAS 294, 283 
\bibitem[1986]{dermer}
Dermer, C. D. 1986,  A\&A 157, 223
\bibitem[1998]{dixon} 
Dixon, D. D. et al., astro-ph 9803237 to appear in New Astronomy 
and Third Int. Symp. on Sources and Detection of Dark Matter in the 
Universe, ed. Cline, D. (Amsterdam: Elsevier), 1998, in press 
\bibitem[1998]{draine}
Draine, B.T., astro-ph 9805083
\bibitem[1994]{fn1}
Fabian, A.C. \& Nulsen, P.E.J. 1994, MNRAS 269, L33
\bibitem[1985]{fall}
Fall, S. M. \& Rees, M. J. 1985, ApJ 298, 18
\bibitem[1987]{fiedler}
Fiedler, R.L., Dennison, B., Johnston, K.J., \& Hewish, A. 1987, 
Nature 326, 675
\bibitem[1990]{gaisser} 
Gaisser, T. K., in {\it Cosmic Rays and Particle Physics}
(Cambridge University Press, Cambridge, 1990)
\bibitem[1996]{gs}
Gerhard, O. E. \& Silk, J. 1996, ApJ 472, 34
\bibitem[1984]{hillas}
Hillas, A. M. 1984, ARAA 22, 425
\bibitem[1990]{kang} 
Kang, H., Shapiro, P. R., Fall, S. M. \& Rees, M. J. 1990, ApJ 363, 488
\bibitem[1997a]{kerins1}
Kerins, E. J. 1997a, A\&A 328, 5
\bibitem[1997b]{kerins2}
Kerins, E. J. 1997b, A\&A 332, 709
\bibitem[1998]{ke}
Kerins, E.J., \& Evans, N.W. 1998, ApJ 503, L75
\bibitem[1997]{mori}
Mori, M. 1997, ApJ 478, 225
\bibitem[1997]{fn2}
Nulsen, P.E.J. \& Fabian, A.C.  1997, MNRAS 291, 425
\bibitem[1987]{sst}
Schlickeiser, R., Sievers, A. \& Thiemann, H. 1987, A\&A 182, 21 
\bibitem[1993]{silk}
Silk, J. 1993, 
Proceedings of Les Houches Summer School on Theoretical Physics on: 
{\it Cosmology and Large-Scale Structures}, eds. Schaeffer, R., 
Silk, J., Spiro, M. \& Zinn-Justin, J., 75 
\bibitem[1998]{simpson}
Simpson, J. A. \& Connell, J. J. 1998, ApJ 497, L85
\bibitem[1998]{sreekumar}
Sreekumar, P. et al. 1998, ApJ 494, 523
\bibitem[1995]{vietri} 
Vietri, M. \& Pesce, E. 1995, ApJ 442, 618
\bibitem[1996]{volk} 
V\"olk, H. J., Aharonian, F. A. \& Breitschwerdt, D. 
1996, Space Sci. Rev. 75, 279
\bibitem[1998]{ww}
Walker, M. \& Wardle, M. 1998, ApJ 498, L125
\bibitem[1995]{wolf} Wdowczyk, J. \& Wolfendale, A. W. 1995,
in {\it 24th International Cosmic Ray Conference}, Vol. 3, 360 
\end{thebibliography}
\end{document}